\documentclass[twocolumn,showpacs,preprintnumbers,amsmath,prb]{revtex4}

\usepackage{graphicx}
\usepackage{dcolumn}
\usepackage{subfigure}
\usepackage{bm}

\graphicspath{{images/}}
\DeclareGraphicsExtensions{.eps,.ps}

\begin{document}


\title{Ground states of two-dimensional $\pm$J Edwards-Anderson spin glasses}

\author{J.W. Landry}
 \email{jwlandry@uchicago.edu}
 \altaffiliation{Present address: Sandia National Laboratory,
   Albuquerque, NM  87185}
\author{S.N. Coppersmith}
 \altaffiliation{Present address: Department of Physics, University of
   Wisconsin, 1150 University Avenue, Madison, WI  53706}
\affiliation{James Franck Institute, University of Chicago,
5640 S. Ellis Ave., Chicago, IL  60637}

\date{\today}

\begin{abstract}  
We present an exact algorithm for finding all the ground states
of the two-dimensional Edwards-Anderson $\pm J$ spin glass
and characterize its performance.
We investigate how the ground
states change with increasing system size and and with
increasing antiferromagnetic bond ratio $x$.
We find that
that some system properties have very large and strongly non-Gaussian
variations between realizations.
\end{abstract}

\pacs{75.50.Lk, 75.10.Nr}
\maketitle

\section{\label{sec:introduction}Introduction}

The Edwards-Anderson (EA) $\pm J$ spin glass~\cite{EdwardsMay1975} is a
canonical example of a system with competing interactions that give rise
to large numbers of low-energy states.  Despite extensive investigation,
the low-energy landscape of this model is still
controversial in both three~\cite{HartmannJan1999,HartmannMar1999,%
PalassiniDec1999,KrzakalaOct2000,KrzakalaMar2001,MarinariMar2000,%
MarinariNov2000,Lamarcq0107544,PalassiniJan2000} and two
dimensions~\cite{SaulNov1993,%
deSimoneJan1996,KawashimaJul1997,ShirakuraOct1997,MatsubaraNov1998,%
HartmannApr1999,ShiomiSep2000,MiddletonAug1999,MiddletonFeb2001,%
PalassiniOct1999,Houdayer0101116,NewmanApr2000,Carter0108050,%
Hartmann0107308}.

Although the two-dimensional $\pm$J Edwards-Anderson model is much
simpler than a three-dimensional case (it does not have a phase
transition at nonzero temperature~\cite{PalassiniApr2001} and individual
ground states can be found in a time that scales as a polynomial of the
system size~\cite{BiecheAug1980}), it still has many metastable states
and a complex energy landscape.  At low temperatures spin glass
relaxation times become very long, complicating investigations using
standard Monte Carlo sampling techniques~\cite{BinderOct1986}, and also
to varying extents more sophisticated sampling methods such as cluster
algorithms~\cite{PerskyJan1996,Houdayer0101116} and multicanonical
methods~\cite{JankeMay1998,HatanoJun2000,BhattacharyaMar1998}.

For ground state properties, exploiting optimization algorithms that
find exact ground states has proven a powerful
approach~\cite{KawashimaJul1997,Rieger9705010,MiddletonAug1999,HartmannFeb2000a}.
However, these algorithms find a single ground state of a single
realization, and one must sample appropriately from the ground states of
each realization~\cite{SandvikJan1999,HartmannMar1999,HartmannFeb2000b}
and also perform a reliable realization average to obtain correct
results.

In this paper we investigate the low-energy properties of the
two-dimensional $\pm$J E-A model by finding exactly {\em all} the ground
states of each realization.  We check that our enumeration is exhaustive
by comparing the number of ground states that are found to exact results
for the partition function obtained using the method of Saul and
Kardar~\cite{SaulNov1993,SaulDec1994}.

Though our algorithm is based on an existing polynomial-time algorithm
that finds individual ground states~\cite{CookMar1999}, it does not run
in polynomial time.  This is impossible because the time just to
enumerate the ground states grows exponentially with system size.
Nonetheless, the number of ground states is vastly smaller than the
number of spin configurations, and empirically we find that our
algorithm runs in a time roughly linear in the number of ground states.
Memory issues limit our current implementation to about $2 \times 10^6$
ground states.  Because there are huge variations in the number of
ground states among realizations, the system sizes that we investigate
are rather small.  Though the median number of ground states of a $10
\times 10$ system in which half the bonds are antiferromagnetic is
$10^4$, at this system size $3$ percent of the realizations have greater than
$2 \times 10^6$ ground states.  Therefore, most of the data presented
here is for systems of size $10 \times 10$ and less.

The advantage of our method is that it produces qualitatively new
information because all the ground states are known explicitly and
exactly, so that one can compute in detail the relationships between
them.  Moreover, these exact results can be used to validate sampling
methods appropriate for larger systems.

We use our method to investigate the paramagnetic-ferromagnetic phase
transition that occurs as $x$, the fraction of antiferromagnetic bonds
in the system, is
increased~\cite{BiecheAug1980,BarahonaMar1982,KawashimaJul1997}.
Quantitative analysis is complicated greatly by
the fact that many quantities exhibit large, non-Gaussian variability
between realizations.  Nonetheless, we are able to obtain new insight
into the nature of the paramagnetic-ferromagnetic phase transition that
occurs as the fraction of antiferromagnetic bonds is increased.

The paper is organized as follows.  Sec.~\ref{sec:method} presents the
algorithm, Sec.~\ref{sec:distribution} presents data on the
distribution of ground states, Sec.~\ref{sec:performance} presents
results on the algorithm performance, and Sec.~\ref{sec:order} describes
our investigation of the destruction of ferromagnetic order as the
fraction of antiferromagnetic bonds is increased.  The results are
discussed in Sec.~\ref{sec:discussion}. The
appendix~\ref{sec:algorithm_details} gives a detailed presentation of
the algorithm.

\section{\label{sec:method}Model and Methods}

\subsection{The Edwards-Anderson model}

We study the two-dimensional Edwards-Anderson (EA)
model~\cite{EdwardsMay1975}, in which nearest-neighbor
Ising spins ($\sigma_i = \pm
1$) on an $L \times L$ square lattice interact either via a ferromagnetic
or an antiferromagnetic coupling.  The Hamiltonian is
\begin{equation}
H = -\sum_{\langle ij \rangle} J_{ij} \sigma_i \sigma_j~,
\label{eq:EAmodel}
\end{equation}
where the sum $\langle ij \rangle$ is over all pairs of nearest-neighbor
spins.  Each bond $J_{ij}$ is chosen to be either $+1$ (ferromagnetic)
or $-1$ (antiferromagnetic).  We designate the fraction of
antiferromagnetic bonds as $x$;
 $x=0$ is the Ising ferromagnet (no disorder), $x=.5$ (with
equal numbers of ferromagnetic and antiferromagnetic bonds) is the
maximally-frustrated spin glass (maximum disorder), and $x=1$ is the
Ising antiferromagnet (no disorder).  Our systems range from $x=.05$ to
$x=.5$ and have periodic boundary conditions.


\subsection{Calculating all the ground states of the EA model}

Our algorithm for finding all the ground states of the EA model first
converts the problem of finding ground states into a graphical matching
problem, as in Refs.~\cite{Toulouse1977,BiecheAug1980}.  Next, all
possible optimal matching solutions of this problem are found, and
finally, these matchings are converted back into spin configurations.

\subsubsection{Conversion of energy minimization to a matching problem}

Refs.~\cite{Toulouse1977,BiecheAug1980} show that the problem of finding
a ground state for this spin glass model can be converted to a matching
problem in graph theory.  Here, we sketch out this conversion and
discuss some subtleties that arise from our use of periodic boundary
conditions.

\begin{figure}
\includegraphics[height=3cm]{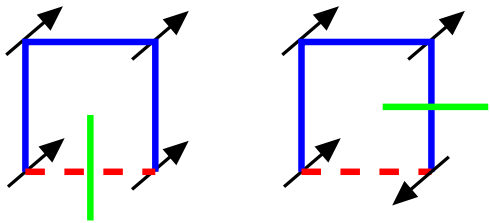}
\caption{Two ground states of a frustrated plaquette with four bonds.
Ferromagnetic bonds are black lines, while antiferromagnetic bonds are
dashed lines. Every configuration of spins produces at least one
unsatisfied bond (denoted with a line perpendicular to the bond), and
there are four minimum energy configurations.}
\label{fig:frustration}
\end{figure}

A ground state of a spin glass can be described not only in terms of
spins and bonds, but also as frustrated plaquettes and paths of broken
bonds~\cite{BiecheAug1980}.  In a frustrated system, it is not possible for
all bonds to be satisfied simultaneously~\cite{Toulouse1977}, which
leads to a natural degeneracy of states.  A simple example is shown in
Figure~\ref{fig:frustration}.

\begin{figure}
\centerline{\includegraphics[width=8cm]{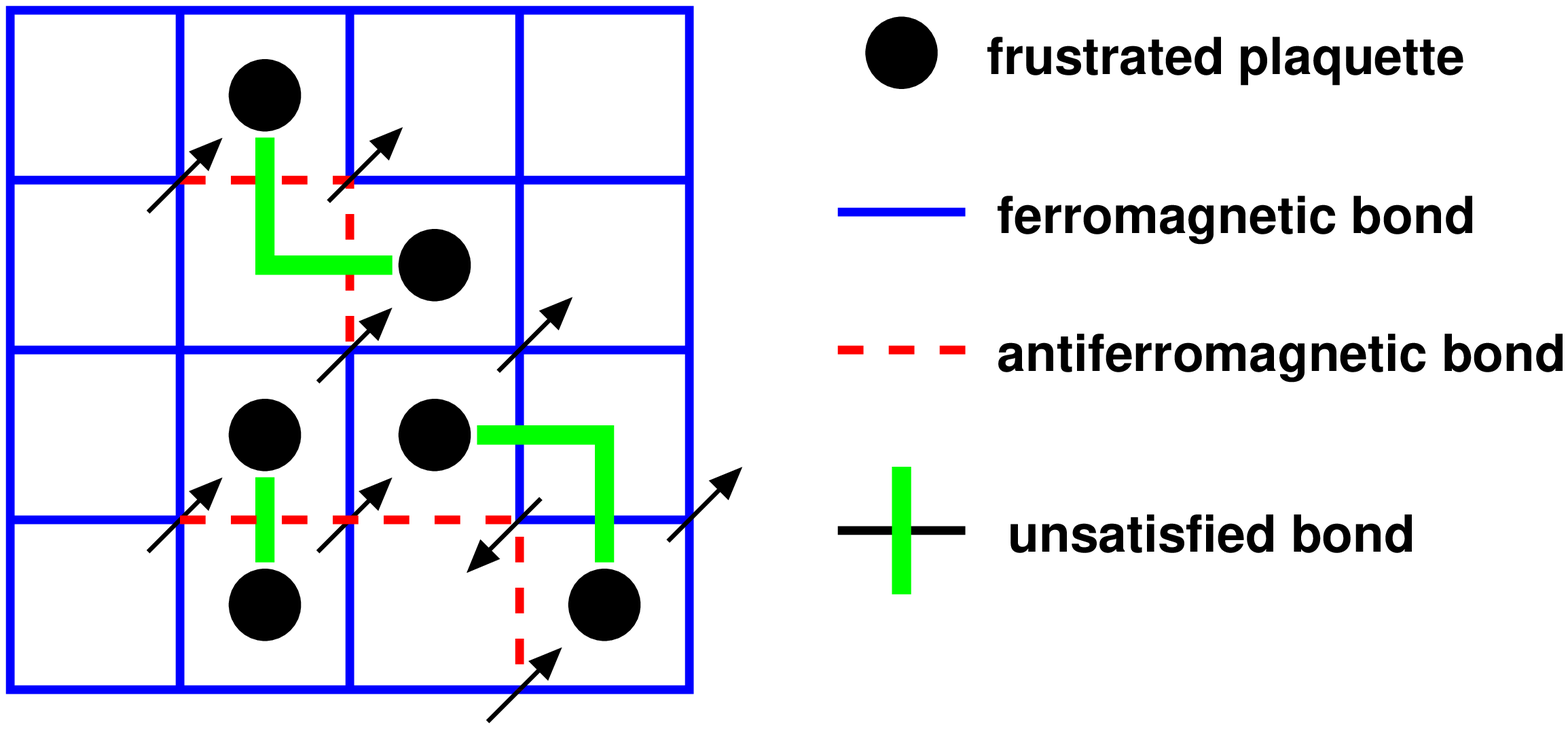}}
\caption{Sample ground state of a spin glass, showing frustrated
plaquettes, unsatisfied bonds, and the corresponding spin and bond configuration.}
\label{fig:groundstate}
\end{figure}

We denote plaquettes with an odd number of unsatisfied bonds as
frustrated, while satisfied plaquettes have an even number of
unsatisfied bonds.  Frustrated bonds form paths that connect frustrated
plaquettes to each other.  Because every frustrated plaquette has an odd
number of frustrated bonds, it must be the endpoint of a path.
Satisfied plaquettes either have no path through them or are midpoints
in a path.  This can be seen in Figure~\ref{fig:groundstate}, where
perpendicular lines have been added to frustrated bonds to show the
paths.

\begin{figure}
\centerline{\includegraphics[width=7cm]{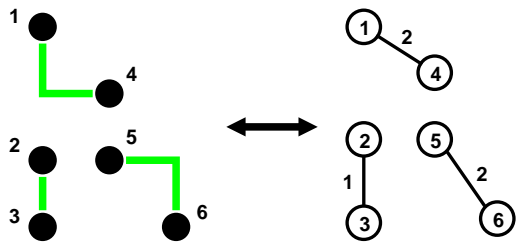}}
\caption{Correspondence of spin glass ground state to solution
of matching problem.  The numbers along each edge of the matching
solution indicate the weight of that edge.}
\label{fig:conversion}
\end{figure}

We identify the frustrated plaquettes as nodes of a graph, and the paths
as edges with a weight equal to the number of broken bonds along the
path.  Ground states have the minimum number of frustrated bonds, so the
problem of finding a spin glass ground state is also the problem of
finding those edges that have the shortest total length.  This problem
arises in the context of graph theory and is called the {\it minimum
weight perfect matching problem}~\cite{CookMar1999}.  In solutions of this
problem, each node is joined to one and only one other node, with the
smallest possible total weight (which corresponds to the lowest possible
energy). Figure~\ref{fig:conversion} illustrates a sample ground state
and its equivalent matching solution.

\subsubsection{Boundary conditions}

Refs.~\cite{BiecheAug1980,Bieche1979} discuss the relation between spin
glass ground states and solutions to a graphical matching problem and
prove a number of results for planar graphs (e.g. free boundary
conditions).  In these references, the ground state problem is first
converted to a matching problem that can be solved in polynomial
time~\cite{EdmondsMar1965,Lawler1976}.  This matching solution is then
shown to correspond always to a spin glass ground state.

For a periodic lattice, the transformation from spins and bonds to nodes
and edges proceeds exactly as in the planar case, and the resulting
matching problem can be solved in polynomial time.  The issue that
distinguishes this problem from the planar case is the conversion of the
matching solution back to a ground state solution.  The matching
solution found will not always correspond to a ground state spin
configuration for a given toroidal boundary condition.  This complication
arises
because four lattices with four different boundary conditions will
produce the same matching problem.  These four boundary conditions are
periodic on all sides, antiperiodic on the top and bottom, antiperiodic
on the left and right, and antiperiodic on all sides.

\begin{figure}
\centering
\subfigure[Periodic NSEW]{\includegraphics[width=1.25in]{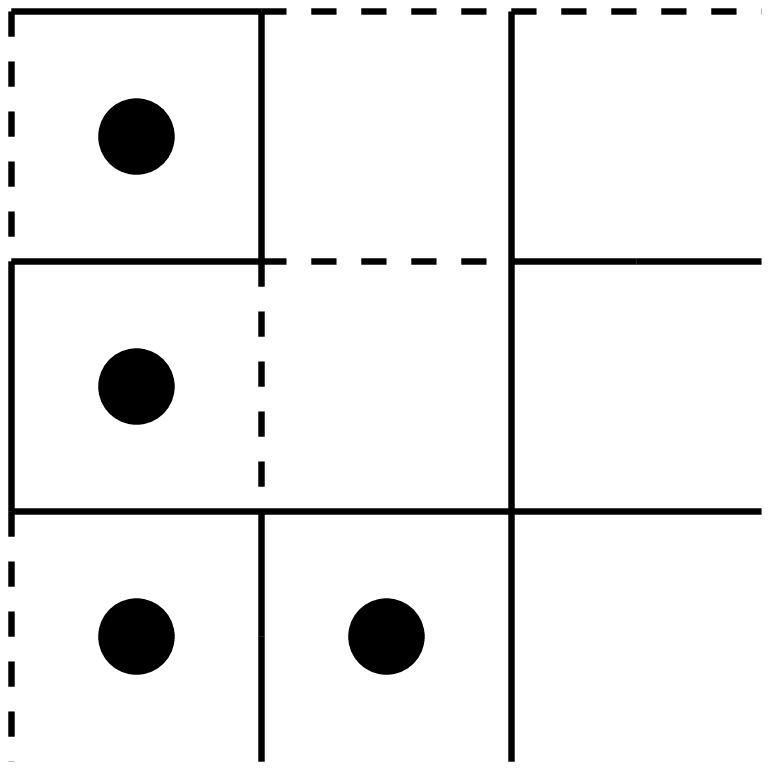}}\quad
\subfigure[Antiperiodic NS, Periodic EW]{\includegraphics[width=1.25in]{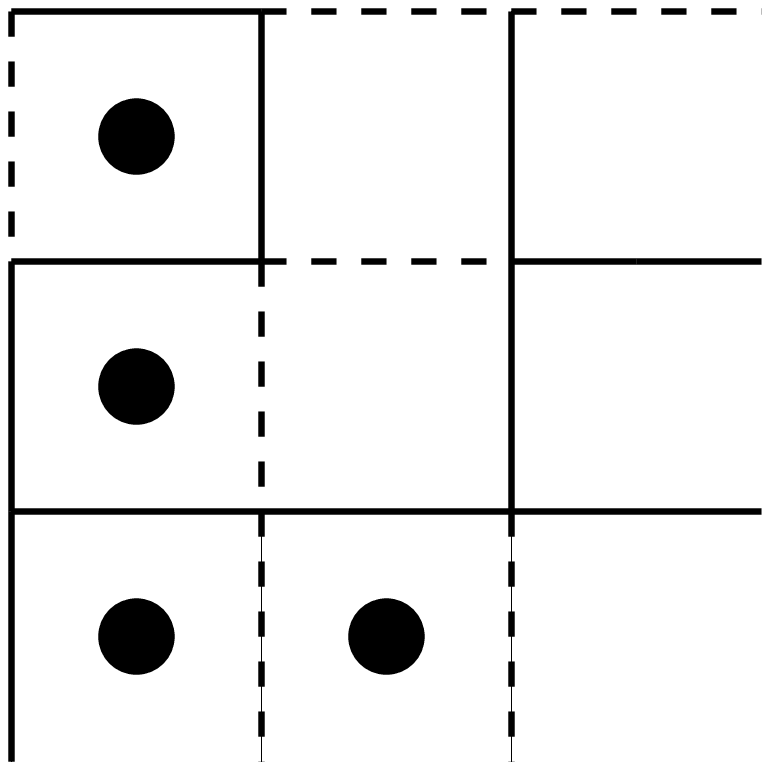}}\quad
\subfigure[Periodic NS, Antiperiodic EW]{\includegraphics[width=1.25in]{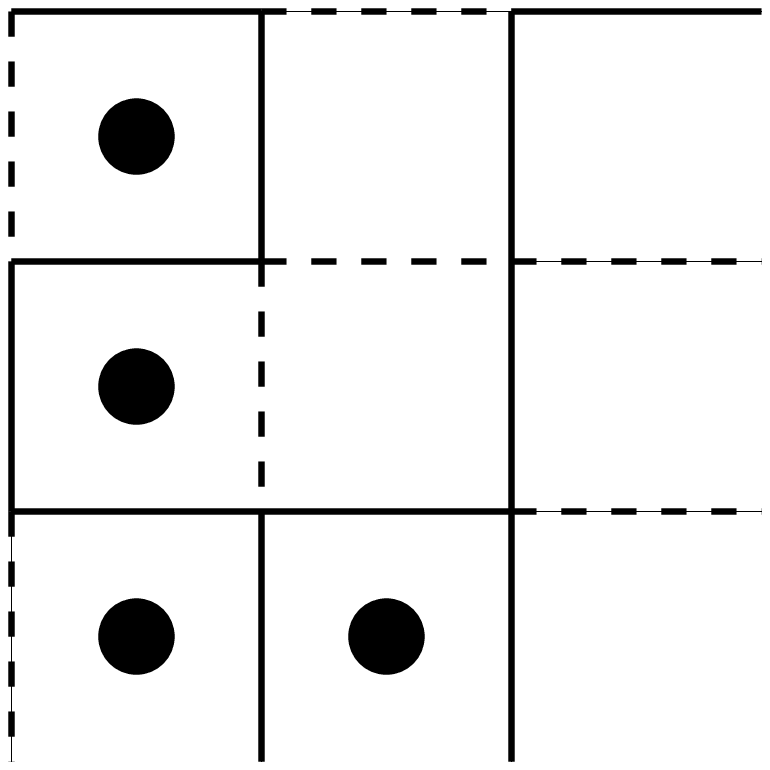}}\quad
\subfigure[Antiperiodic NSEW]{\includegraphics[width=1.25in]{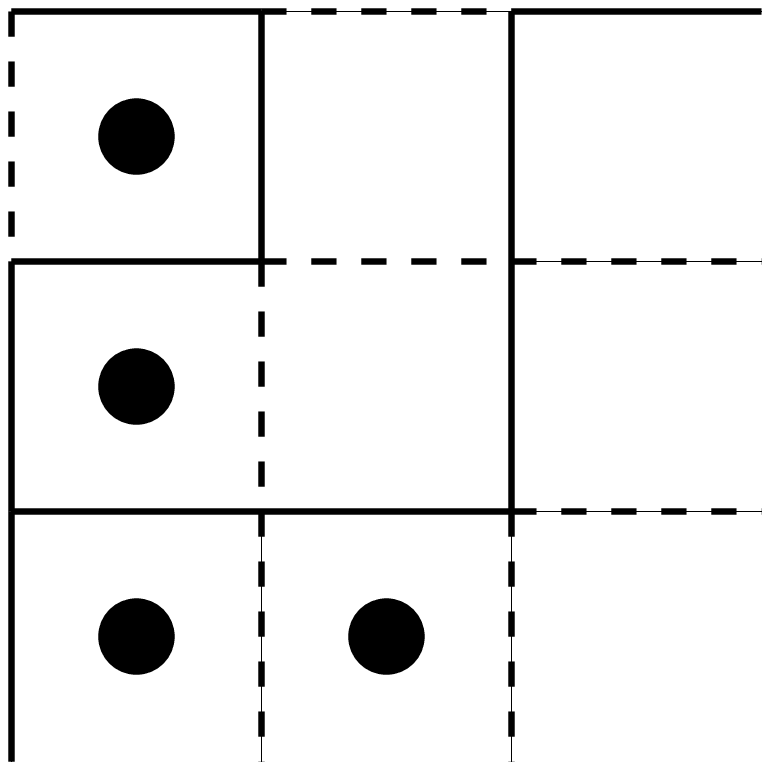}}
\caption{The four different boundary conditions for a sample lattice.
Ferromagnetic bonds are solid lines and antiferromagnetic bonds are
dashed lines.  Bonds on the right and bottom sides wrap around the
lattice to reconnect on the other side.  Each boundary
condition has the same set of frustrated plaquettes, shown as filled-in
circles.}
\label{fig:latticebcs}
\end{figure}

Boundary conditions can be changed from periodic to antiperiodic on an
$L \times L$ system either by setting $J^{new}_{iL} = - J_{iL}$
along the desired edge or by flipping the spins so that $S^{new}_{i1} =
-S_{i1}$ along the desired edge.  In this study, we flip the bonds.  A
sample lattice with the four different boundary conditions is shown in
Figure~\ref{fig:latticebcs}.

These four lattices have exactly the same frustrated plaquettes, so they
produce the same matching problem.  In this sense, the matching problem
does not understand boundary conditions.  When a matching solution is
converted back into spins and bonds, it may not correspond to a ground
state for a given boundary condition.

We resolve this ambiguity by converting explicitly each matching into
a spin configuration and checking the viability of each spin
configuration for each boundary condition.  A ground state is only
accepted for a given boundary condition if it has a consistent spin
configuration.  We find numerically that a matching solution always
corresponds to a ground state solution of at least one boundary
condition.

This subsubsection has described the necessary procedures for generating
all the ground states in the case of periodic boundary conditions.  The
algorithm works perfectly for planar graphs without these procedures.

\subsubsection{Generating all optimal matchings}

Our algorithm for finding all the optimal matchings has three parts.
The first part finds all edges that make up the optimal solutions.  This
part exploits the structure of the edges, since the number of edges
appearing in the ground states is a small subset of the total number of
edges. The second part takes this subset of edges and combines them to
find all optimal matchings.  The third part converts the optimal
matchings into ground state configurations.  The next section describes
our algorithm briefly.  More detail, including an example, is in the
appendix~\ref{sec:algorithm_details}.

The algorithm uses Edmonds' blossom
algorithm~\cite{EdmondsMar1965,Lawler1976}, which finds a single optimal
solution to a matching problem in polynomial time.  We use the Concorde
implementation of the algorithm~\cite{CookMar1999,ConcordeWebsite}.

\subsection{Finding all edges in all solutions}

The algorithm begins by making a list of nodes and possible edges.  All
frustrated plaquettes are found and designated as nodes.  Pairs of nodes
that are within a distance of five are considered to have edges between
them.  This restriction controls the combinatorial explosion of possible
edges, and optimal solutions involving weights larger than five are
incredibly rare.  Moreover, any errors introduced by
this truncation are identified and eliminated at a later stage when the
total number of ground states found for a realization is compared to an
independent determination of the ground state degeneracy.

To construct a list of edges that exist in at least one minimal weight
matching, which we designate as viable edges, we start with an empty
list.  The blossom algorithm is run on the unmodified matching problem
and a single matching solution is found.  Each edge in this solution is
added to the list of viable edges.

To find more viable edges, the nodes are considered successively.  For
each node, a modified list of edges is created from the original list by
deleting those known viable edges connecting to the current node.  The
blossom algorithm is run on this modified list to find an optimal
solution for this new problem.  If the solution has the correct path
length (i.e. corresponds to a ground state), then the new viable edges
that have been found are added to the list of viable edges.  The process
continues for this node.  If the path length of the new solution is too
large (i.e. does not correspond to a ground state), then we know that we
have found all the viable edges associated with this node.  The
algorithm then proceeds to the next node in the list.  By moving
sequentially through the nodes, all viable edges are found.

\subsection{Determining optimal matchings}

The next part of the algorithm uses the list of viable edges to find all
of the optimal matchings.  It picks edges systematically from the list
of viable edges until each node is connected by a given edge to one and
only one other node.  All possible combinations of viable edges in which
each node is incident on exactly one edge are examined.

Whenever there is this kind of combination of elements, there is a
danger of a combinatorial explosion.  In this case, the number of
matchings (combinations of edges) is relatively controlled.
Sec.~\ref{sec:performance} discusses this issue in detail.

\subsection{Converting matchings to ground states}

\begin{figure}
\centerline{\includegraphics[width=6cm]{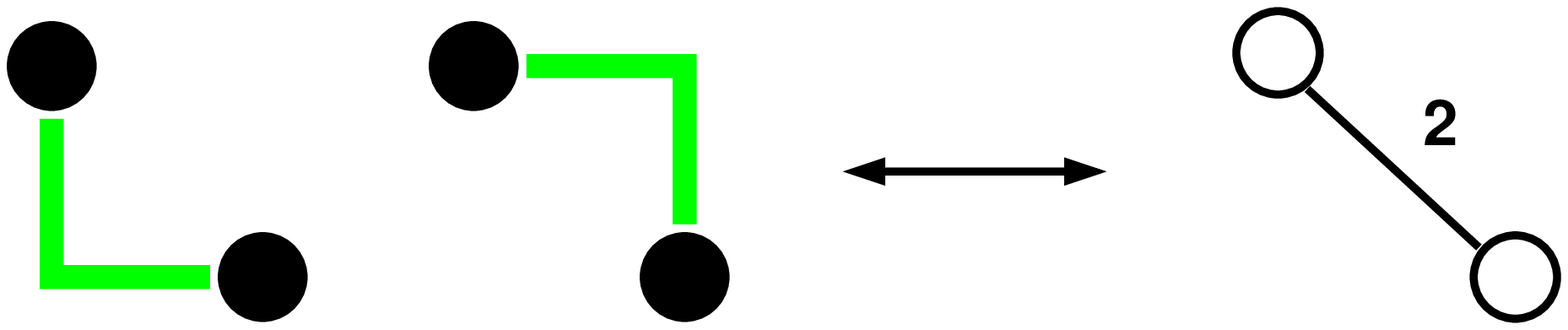}}
\vskip .2in
\centerline{\includegraphics[width=7cm]{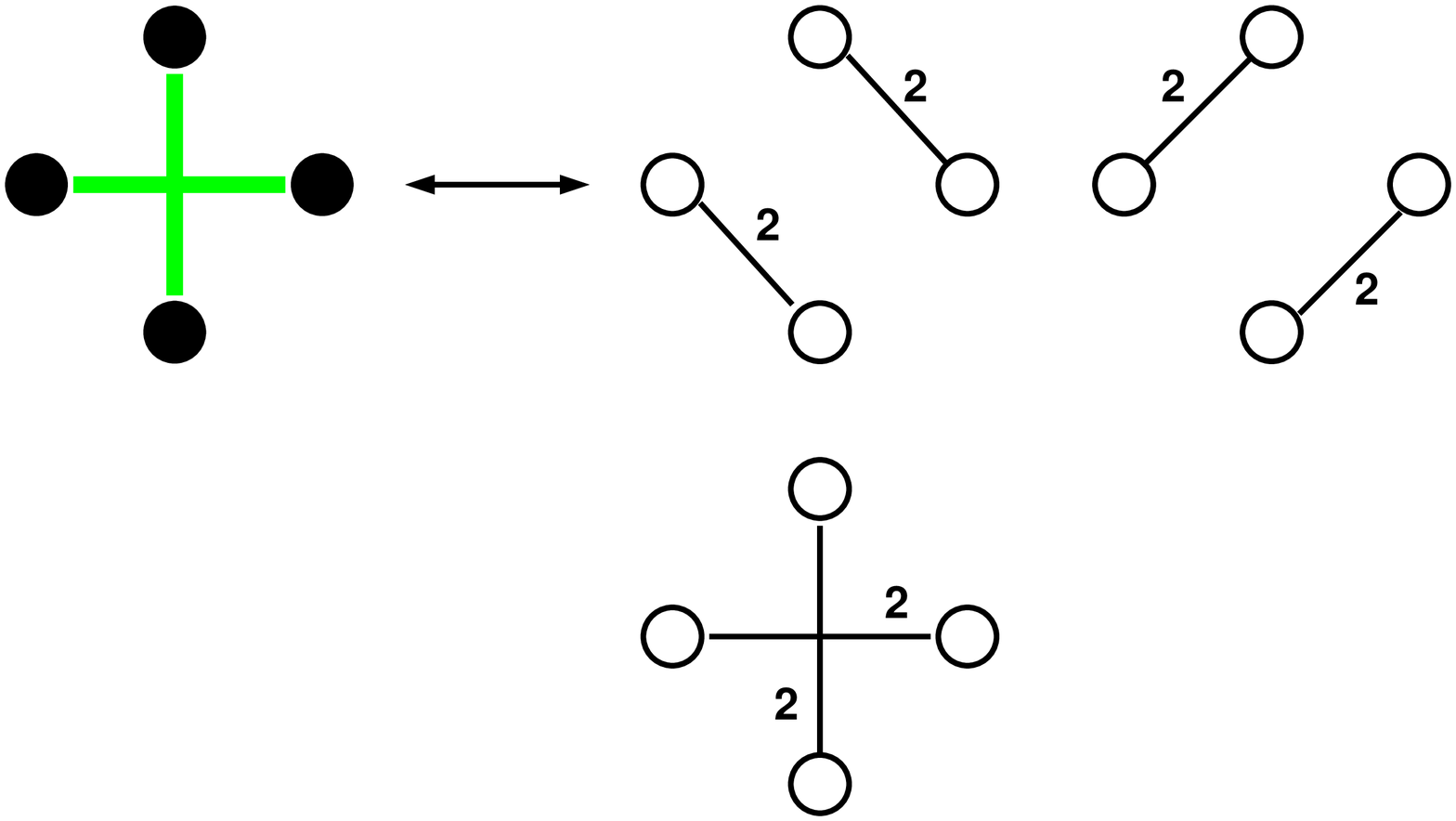}}
\caption{The relationship between matching solution is not one-to-one.
As shown in the top diagram, two or more different ground states can
correspond to the same matching solution.  In addition, as seen in the
bottom diagram, a single ground state can correspond to multiple
matching solutions.  The filled circles are frustrated plaquettes, the
thick dark lines are unsatisfied bonds, the open circles are nodes, and
the thin dark lines are edges with lengths as shown.}
\label{fig:matchtoGS}
\end{figure}

All optimal solutions to the matching problem must be converted back
into ground state spin configurations.  This conversion is nontrivial
because one matching solution can correspond to many different ground
states, and the same ground state can be represented by different
matchings.  Simple examples of this phenomenon can be seen in
Figure~\ref{fig:matchtoGS}.  To resolve these complications, we keep
every ground state we find in memory.  Any proposed ground state is
checked to see that it does not correspond to a ground state we have
already found.

The other important issue is the role of boundary conditions in this
conversion from matchings to ground states.  We determine the ground
state or ground states from the matching by fixing the value of a single
spin (in our case, we fix the upper left-hand spin as $+1$).  Every
other spin follows from this initial spin, because we know the specific
bonds of the current boundary condition and their status as satisfied or
unsatisfied.

\subsection{Partition function}

To check that our algorithm finds every ground state, we also generate
the partition function of the realization at $T=0$.  This partition
function gives the number and energy of the ground states of a given
realization.  We generate a partition function in polynomial time using
Saul and Kardar's technique,~\cite{SaulNov1993,SaulDec1994} which is a
generalization of methods used for finding the partition function at
$T=0$ for the two-dimensional Ising
model.~\cite{KacDec1952,PottsJan1955,BurgoyneOct1963,GlasserAug1970} For
reasons of computational efficiency, we consider only $L \times L$
lattices where $L$ is even.  Because our methods yield ground states not
only for lattices with regular periodic boundary conditions but also for
those with antiperiodic boundary conditions, we generate four different
partition functions for each possible lattice, corresponding to the four
different boundary conditions mentioned above.

We are confident that our algorithm works properly because the number of
ground states found by our algorithm agrees with the partition function
result for every realization and boundary condition that we have
examined.

\section{\label{sec:distribution}Ground State Distribution}
Before presenting the results from our algorithm, we first discuss the
distribution of numbers of ground states for different realizations with
varying system size $L$ and antiferromagnetic bond ratio $x$.  All of
these results were obtained from the partition functions of these
realizations, generated using Saul and Kardar's
method~\cite{SaulDec1994}.  These ground state distributions show large
sample-to-sample variations for realizations with the same $L$ and
$x$.

\begin{figure*}
\centerline{\includegraphics[height=15cm,angle=270]%
{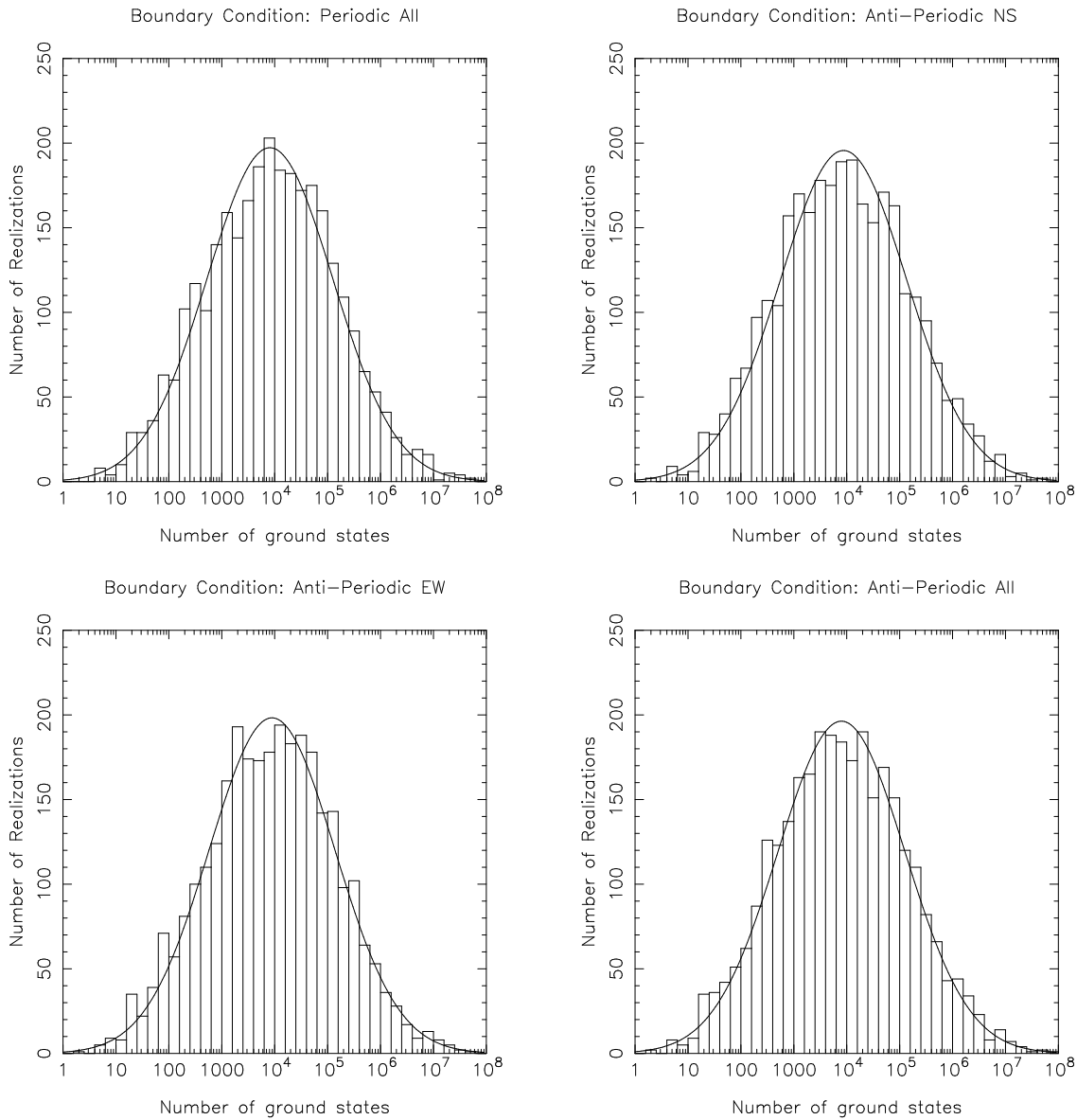}}
\caption{Histograms of the number of ground states for 3006 realizations
with $L=10$ and $x=.5$ for the four different boundary conditions.  The
solid lines are $\chi^2$ fits to
the log-normal distribution, Eq.~\ref{eq:log_normal},
with parameters are given in Table~\ref{tbl:gs-10-5-bc}.
At this value of $x$, changing boundary conditions has very little effect on the
distribution of ground states.}
\label{fig:gs-10-5-bc}
\end{figure*}

Figure~\ref{fig:gs-10-5-bc} shows four histograms of the number of
ground states for the four different boundary conditions of 3006
realizations of $L=10$ lattices with $x=.5$.  The solid lines on the
figure are fits to a log-normal
distribution~\cite{Aitchison1957,Crow1988,LimpertMay2001}, where the
number of realizations with between $m$ and
$m +dm$ ground states is $G(m;\mu,\sigma)d(\log_{10}(m))$, with
\begin{equation}
G(m;\mu,\sigma) = \frac{A}{\sigma\sqrt{2\pi}}
\exp \left[\frac{-(\log_{10}(m)-\mu)^2}{2\sigma^2} \right]
~.
\label{eq:log_normal}
\end{equation}
Here, $\mu$ is the most probable value of $\log_{10}(m)$, $\sigma$
describes the width of the distribution, and $A$ is a normalization
constant.  Table~\ref{tbl:gs-10-5-bc} gives the parameter values
from a $\chi^2$ fit with the errors for the bin heights taken to be
$\sqrt{N_b}$, where $N_b$ is the number of realizations in a given bin.
The distributions fit a log-normal distribution extremely well, and all
the parameters of the fit for the four different boundary condition are
consistent with each other within error bars.

\begin{table}
\centering
\begin{tabular}{|c|c|c|c|} \hline
Boundary Condition & $\mu$ & $\sigma$ & $A$ \\ \hline
Periodic All & $3.91 \pm .066$ & $1.191 \pm .047$ & $589 \pm 32$ \\ 
Anti-Periodic NS & $3.94 \pm .065$ & $1.199 \pm .044$ & $588 \pm 31$ \\ 
Anti-Periodic EW & $3.95 \pm .056$ & $1.189 \pm .040$ & $591 \pm 28$ \\ 
Anti-Periodic All & $3.90 \pm .057$ & $1.201 \pm .040$ & $591 \pm 28$ \\
\hline
\end{tabular}
\caption{Fits of the boundary condition distributions to a Gaussian with
mean $\mu$, standard deviation $\sigma$, and amplitude $A$.  $\mu$ and
$\sigma$ are given in terms of $z$, the $\log_{10}$ of the number of
ground states.  Boundary condition has little effect on the distribution on
ground states.}
\label{tbl:gs-10-5-bc}
\end{table}

This log-normal distribution means that the variations in the number of
ground states of different realizations are enormous.  Sampling a
few realizations will not give a meaningful picture of the behavior of
the system.  Averages over realizations need significant numbers to
produce reasonable results, and still may not give sufficient
information.  Also, because the distribution of ground states is so
wide, our methods to find ground states and apply perturbations to them
have wildly varying performance on realizations with the same $L$ and
$x$.  A few outliers with many ground states will completely dominate
the computation time of all algorithms.  A change in thinking is
necessary -- the concept of an average realization or number of ground
states is not necessary useful in considering the behavior of this
system.

Figure~\ref{fig:gs-10-5-bc} also demonstrates that at $x=0.5$ the
boundary condition does not affect the ground state distribution.
All future results in this section will be presented for
the fully periodic boundary condition.

\begin{figure}
\centerline{\includegraphics[width=6cm,angle=270]{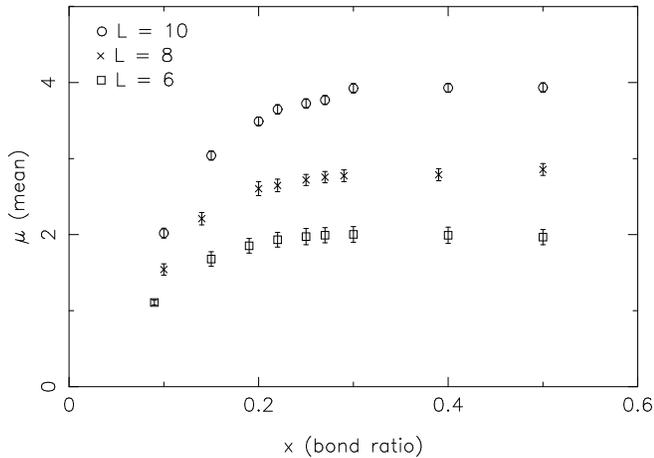}}
\caption{Plot of the parameter $\mu$ defined in Eq.~\ref{eq:log_normal}
(which is the base-10 logarithm of the maximum
of the ground state distribution), versus $x$, the
fraction of antiferromagnetic bonds.  Data for systems with $L=6,8,10$
are shown; the qualitative features do not exhibit strong
$L$-dependence, except for the overall exponential scaling of the number
of ground states with system size.}
\label{fig:saturationmu}
\end{figure}

\begin{figure}
\centerline{\includegraphics[width=6cm,angle=270]{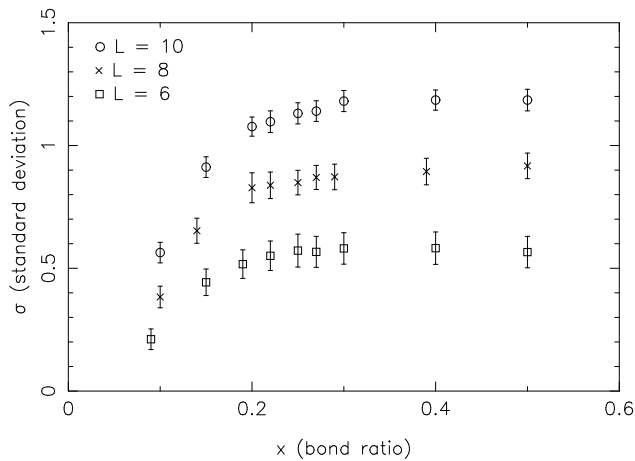}}
\caption{Plot of the parameter $\sigma$ defined in
Eq.~\ref{eq:log_normal}, which is a measure of the width of the
log-normal distribution of
ground states, versus $x$, the fraction of antiferromagnetic bonds.
Data for systems with $L=6,8,10$ are shown; the qualitative features do
not exhibit strong $L$-dependence, except for the overall exponential
scaling of the number of ground states with system size.}
\label{fig:saturationsig}
\end{figure}

Next we study how the distribution of ground states varies with $x$.
Figures~\ref{fig:saturationmu} and~\ref{fig:saturationsig} show how the
parameters $\mu$ and $\sigma$ characterizing the mean and the width of
the distribution change as $x$ is varied between $0$ and $0.5$.  Both
the mean and the width tend to increase with $x$ until they saturate
between $x=.25$ and $x=.3$.

The saturation of the ground state distribution at $x=.3$
appears to be completely distinct from the breakdown of ferromagnetic
order at $x \approx .1$, and seems to be relatively insensitive to
changes in system size.  Since the distribution of ground states is
essentially unchanged from $x=.3$ to $x=.5$, this suggests that systems
in this range of parameters have no essential physical differences.

\begin{figure}
\centerline{\includegraphics[width=11cm,angle=270]{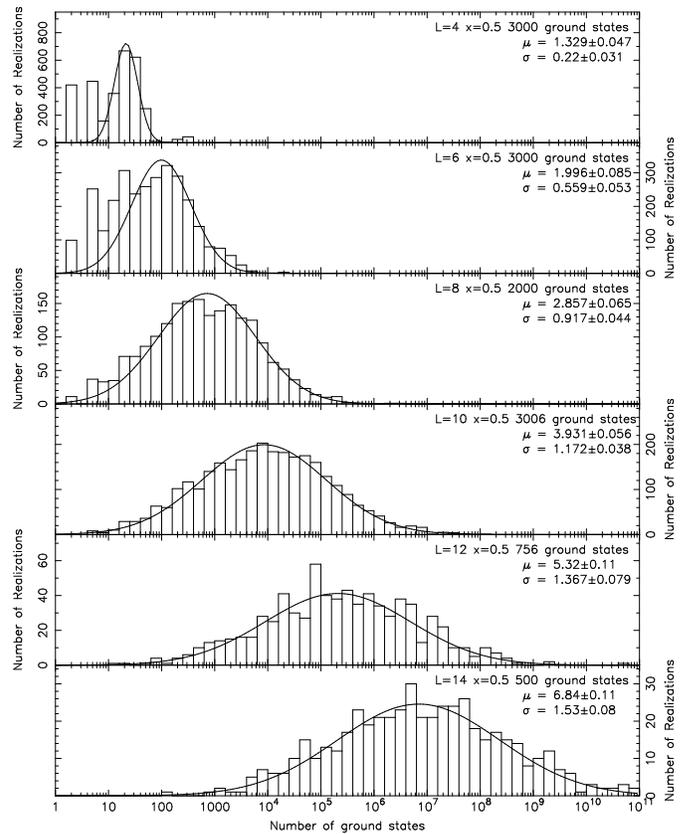}}
\caption{Histograms of the number of ground states for realizations of
different system sizes $L$ = 4, 6, 8, 10, 12, and 14 at $x=0.5$.  The
solid lines are $\chi^2$ fits to log normal distributions.
The parameters $\mu$ (which describes the mean) and $\sigma$
(which describes the width) are defined in Eq.~\ref{eq:log_normal};
they are given
in terms the base-10 logarithm of the number of ground states.
Increasing system size increases both $\mu$
and $\sigma$.}
\label{fig:gs-L-5}
\end{figure}

\begin{figure}
\centerline{\includegraphics[width=7cm,angle=270]{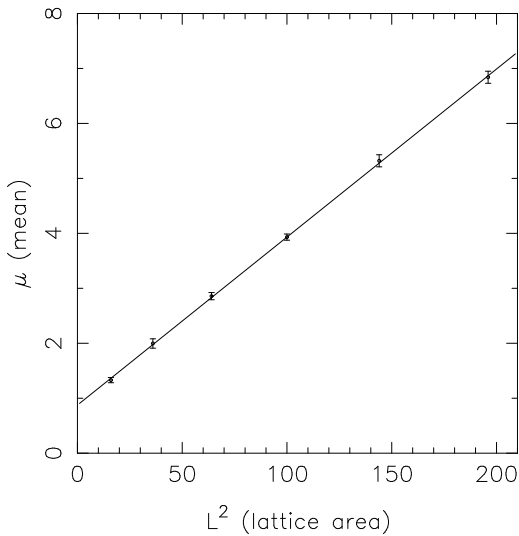}}
\caption{The parameter $\mu$ of the ground state distribution,
defined in Eq.~\ref{eq:log_normal}, versus system
area $L^2$ for systems with $x=0.5$.
The solid line on the graph is a fit to the form $\mu =
(a + bL^2)$, with $a = .874 \pm .025$ and $b=.03058 \pm .00023$.
The most probable number of ground states, $10^\mu$,
increases exponentially with $L^2$.
}
\label{fig:meanscaling}
\end{figure}

Finally, we present the variation of the distribution of ground states
with $L$ at $x=0.5$.  As Figure~\ref{fig:gs-L-5} shows, increasing the system size
moves the ground state distribution over to larger numbers of ground
states but does not change the log-normal distribution of the states.
Again we fit these distributions to the form of Eq.~\ref{eq:log_normal}.
As seen in Fig.~\ref{fig:meanscaling}, the mean $\mu$ of the ground
state distributions at $x=0.5$ scales exponentially with lattice area $L^2$; $\mu
\propto e^{bL^2}$, with $b \approx 0.03$.
It is because $b \ll \ln 2$ that our algorithm finds all the ground states much
more efficiently than an exhaustive search of all configurations.

\begin{figure}[floatfix]
\centerline{\includegraphics[width=7cm,angle=270]{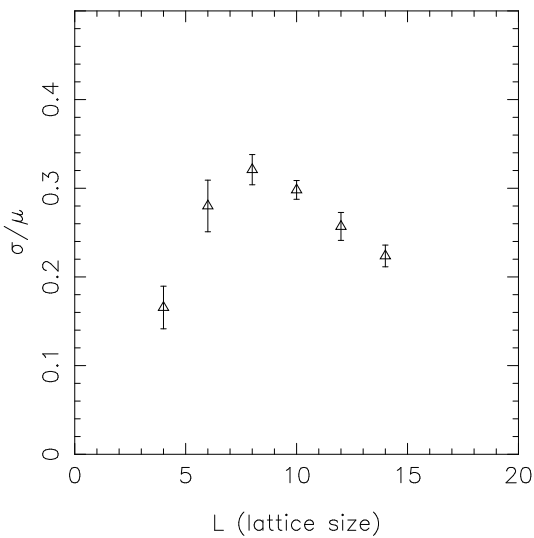}}
\caption{Plot of the ratio $\sigma/\mu$ versus $L$.  This is a test of
self-averaging for the free energy.  Note that $\sigma/\mu$ increases
initially and then begins to fall off as $L$ increases in a roughly
linear way.}
\label{fig:selfaveraging}
\end{figure}

We also investigate whether the distribution of ground states is self
averaging~\cite{DerridaSep1997}, that is, whether the ratio of the width
of the distribution to its mean vanishes in the thermodynamic limit $L
\rightarrow \infty$.  Figure~\ref{fig:selfaveraging}, a plot of
$\sigma/\mu$ as a function of $L$, shows that $\sigma/\mu$ actually
increases with $L$ up to $L=8$ (a cautionary note for higher dimensions,
where studies of $L>8$ are extremely difficult), but that $\sigma/\mu$
decreases as $L$ increases for $L>8$, consistent with self-averaging
behavior for the limit $L \rightarrow \infty$.  Data for larger lattices
would be extremely helpful in determining the behavior of $\sigma/\mu$
as system size increases.

The wide distribution in numbers of ground states can be rationalized in
terms of the matching problem by noting that
random arrangements of bonds produce
relatively random arrangements of frustrated plaquettes.
If we treat the number of possible edge choices for different
plaquettes as random variables, and assume that they
are essentially independent from plaquette to plaquette,
then the number of total possibilities is multiplicative,
and it is well-known that multiplicative random processes lead to
log-normal distributions~\cite{Aitchison1957,Crow1988,LimpertMay2001}.
Moreover, since
entropy is the logarithm of the number of accessible configurations, it
is natural to expect $\log(m)$ to be normally distributed.
However, though these simple considerations make the log-normal
form plausible, they do not address the distribution width.

\section{\label{sec:performance}Ground state algorithm performance}

In this section we present the performance of our ground state algorithm
and discuss the steps limiting its performance.  One typically
investigates the performance of an algorithm on a finite lattice by
showing its scaling behavior relative to system size~\cite{SaulDec1994}.
In this instance, such an approach does not provide much understanding
because of the log-normal distribution of ground states for a given
system size and $x$.  Any average over such a broad distribution would
not convey much information.

We instead study our algorithm's performance relative to the number of
ground states, which we believe is a much better measure of performance.
In addition, because the number of ground states of a realization can be
generated quickly from the partition function~\cite{SaulDec1994}, this
measure gives a useful predictor of time to completion before the
algorithm is run.

For all of the following results, the algorithm was run on a Pentium III
800 MHz machine, with 512 MB of RAM\footnote{The machine was a part of a
Beowulf Linux cluster (details available online~\cite{Honeycomb}).
Since the code was not parallelized, each run was only on a single
drone.}.

\begin{figure}[floatfix]
\centerline{\includegraphics[width=6cm,angle=270]{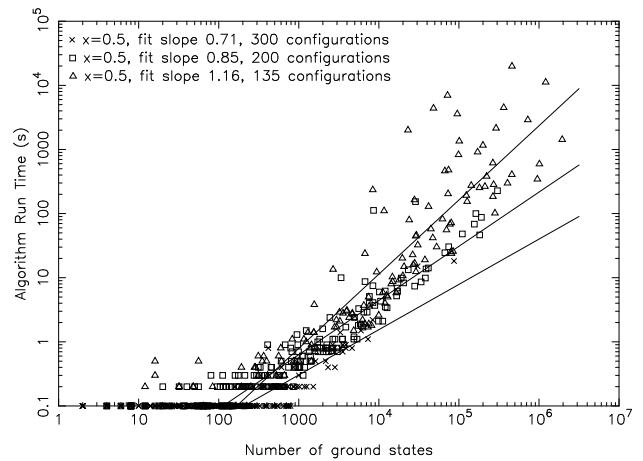}}
\caption{Algorithm run time versus the number of ground
states.  The fits are power law fits to each of the separate data sets
with more than 100 ground states.}
\label{fig:alg-cputime1}
\end{figure}

Figure~\ref{fig:alg-cputime1} is a plot of the algorithm's run time
versus number of ground states.  The plot also shows power law fits of
the results for different system sizes.
Though the scatter is
substantial, it can be seen from the plot that the run times
remain reasonably well-behaved as the number of ground states increases.
Assuming enough memory is available, we expect that systems with $10^8$
ground states could be completed in a week of running time.

\begin{figure}
\centerline{\includegraphics[width=6cm,angle=270]{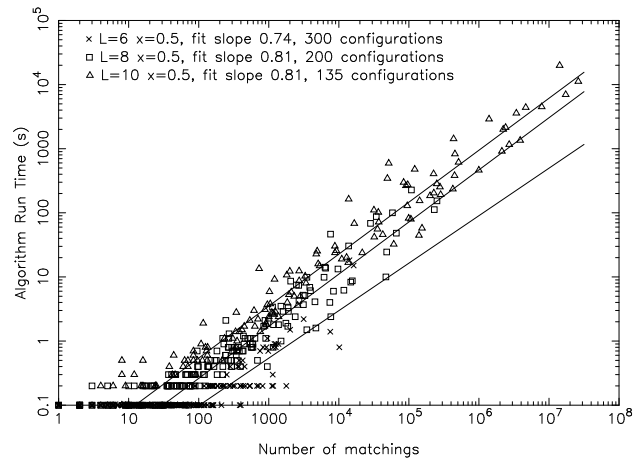}}
\caption{Algorithm run time versus the number of matchings.
The power law fits yield exponents that agree and give a scaling of
$O(n^{4/5})$.  Lattices with less than 100 matchings were excluded from
the fit because of start-up time costs.}
\label{fig:alg-cputime2}
\end{figure}

Figure~\ref{fig:alg-cputime2} demonstrates a more useful way to
characterize the scaling behavior of our algorithm.  Here we plot the
run time versus the number of matchings (sets of edge combinations)
explored.  Results from different system sizes agree and scale as
$\approx O(n_{m}^{4/5})$, where $n_m$ is the number of matchings.  This
measure of system performance is not a very good predictor, because it
is difficult to determine the number of matchings before running the first
part of the algorithm.  In the fits for both
Figures~\ref{fig:alg-cputime1} and~\ref{fig:alg-cputime2}, runs with
very small numbers of matches and ground states are not included,
because there is an initial overhead in computation irrespective of the
scale of the problem.

\begin{figure}
\centerline{\includegraphics[width=6cm,angle=270]{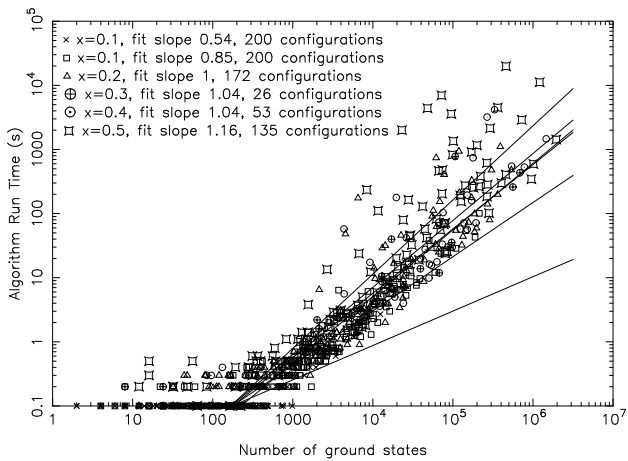}}
\caption{Algorithm run time as a function of the number of ground states
for $L=10$ realizations with various $x$ values.  The power law
fits are to each of the separate data sets.}
\label{fig:alg-cputime3}
\end{figure}

Figures~\ref{fig:alg-cputime1} and~\ref{fig:alg-cputime2} show data for
different lattice sizes $L$ with the same $x=0.5$.
Figure~\ref{fig:alg-cputime3} demonstrates that the same scaling of
algorithm run time with ground states is observed if we vary $x$ and
keep $L$ fixed.

The algorithm scales as a power law with number of matchings, because
the core of the algorithm (its second part) runs sequentially through
all possible matchings of edges in ground states.  We thus expect it to
scale as $O(n_m)$, where $n_m$ is the number of matchings.

The algorithm's scaling with ground states is more complicated,
precisely because the algorithm deals with matchings instead of ground
states, except when inserting into the list of ground states, which
should go like $O(n_{gsbc}^{3/2})$, where $n_{gsbc}$ is the number of
ground states for a given boundary condition.  In practice, the number
of ground states is smaller than the number of matchings by about an
order of magnitude.  Also, since we observe that ground states are split
relatively evenly among the boundary conditions in realizations with
large numbers of ground states, we expect $n_{gsbc}$ to be about one and
a half orders of magnitude smaller than $n_m$.  Since significantly more
work is done only on the ground state matchings, which require an
additional investment of searches, the interplay between the time for
matchings and for the ground state conversions is nontrivial.

\begin{figure}
\centerline{\includegraphics[width=6cm,angle=270]{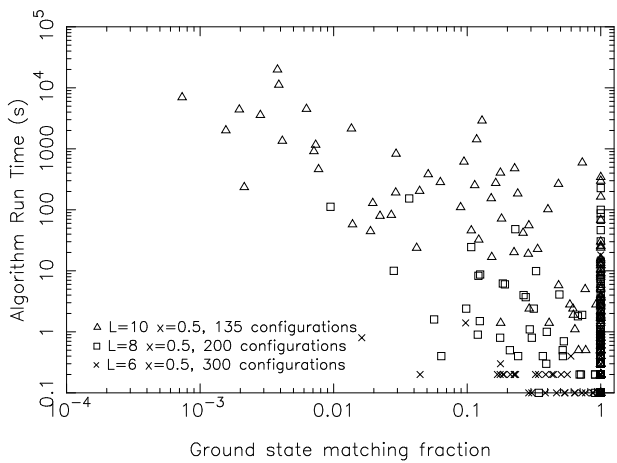}}
\caption{Algorithm run time versus the fraction of all matchings
constructed from viable edges that are ground state matchings.
Small ratios imply long run times.}
\label{fig:alg-matchings}
\end{figure}

The algorithm performs well only if the ground state matchings are a
significant fraction of all possible combinations of viable edges.  If
two distinct systems have the same number of ground states, but one has
10 times as many matchings as the other, their run time will be very
different.  We denote those matchings that lead to a ground state as
ground state matchings, and the ratio of ground state matchings to
number of total matchings of viable edges is inversely correlated with
the run time of the algorithm.  This is shown in
Figure~\ref{fig:alg-matchings}, a plot of run time versus ground state
matching ratio.  Realizations with small ground state matching ratios
have large run times, because the algorithm spends most of its time
cycling through matchings that do not yield ground states.

\begin{figure}
\centerline{\includegraphics[width=11cm,angle=270]{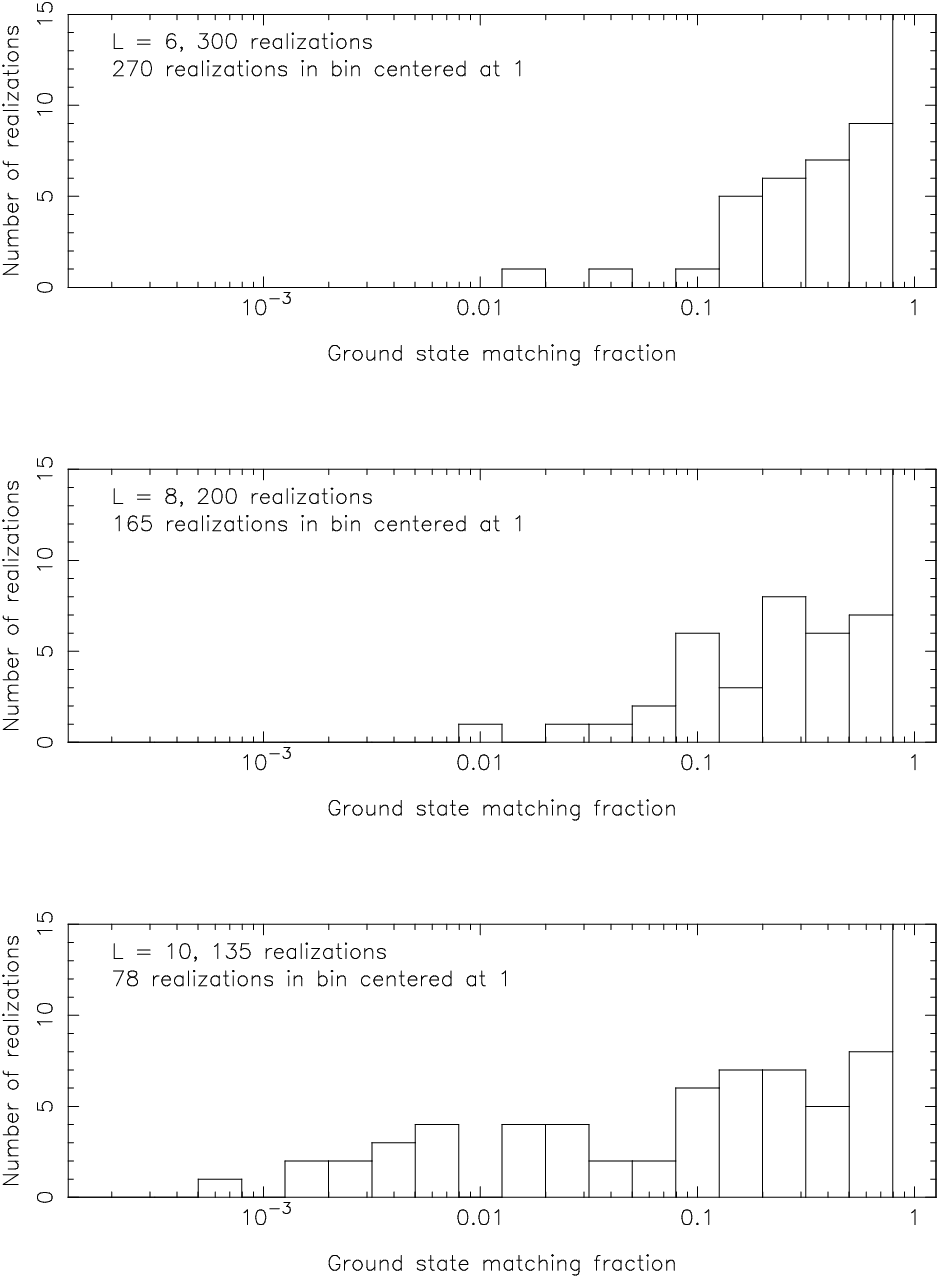}}
\caption{Histograms of the ground state matching ratios for realizations
with $L=6$, 8, and 10.  In all cases, at least half of the realizations
had ground state matching fractions equal to 1.  Note the decreasing
ground state matching fractions as system size increases.}
\label{fig:gsmatchings-hist}
\end{figure}

To illustrate the behavior of the ground state matchings, we present in
Figure~\ref{fig:gsmatchings-hist} a histogram of the ground state
matching ratios for different lattice sizes $L$.  Realizations with
small ground state matching ratios occur rarely, but increase in
frequency with increasing system size.  However, even for $L=10$, over
78 percent of the configurations have matching ratios greater than
$0.1$.
We thus expect this ratio to remain
appreciable for somewhat larger system sizes.

The above plots do not show ground state numbers higher than $2 \times
10^6$, because of memory limitations of our hardware.  The code
currently stores all of the ground states while running to prevent
duplication of ground states.  With more memory, much larger numbers of
ground states could be investigated relatively quickly with this
algorithm, as seen in the scaling above.  In addition, if the algorithm
could be modified so that it did not require all the ground states to be
stored, this memory constraint would be removed.  This would require
some cleverness about how matchings are converted into ground states.
One could conceive of keeping only a list of ``problematic'' ground
states that correspond to multiple matchings, if they could be
determined easily.

\section{\label{sec:order}Destruction of Ferromagnetic Order}
In this section we investigate the destruction of ferromagnetic
order that occurs as $x$, the fraction of antiferromagnetic bonds,
is increased~\cite{KirkpatrickNov1977,GrinsteinJan1979,%
Vannimenus1979,BarahonaMar1982,%
KawashimaJul1997,Bendisch1997,Bendisch1998,Achilles2000}.

As Bendisch and collaborators
discuss~\cite{Bendisch1997,Bendisch1998,Achilles2000}, investigating how
the ground state energy depends on boundary condition is a powerful
method for locating the transition at which ferromagnetic order is
destroyed.  In a system of infinite size, when $x$ is less than the
transition point $x_c \approx 0.1$, one expects all the lowest energy
states to occur when the boundary conditions are consistent with
ferromagnetic order, while for $x > x_c$, there should be no preference
for this type of boundary condition.
Refs.~\cite{Bendisch1997,Bendisch1998,Achilles2000} present calculations
supporting this picture for different non-toroidal boundary conditions.
Our results obtained by computing the partition function for toroidal
boundary conditions also support this picture.  However, because of our
relatively small system sizes, finite size effects in our calculations
are large.  We did not do a finite-size scaling analysis of our data
because we do not expect it to yield qualitatively new information about
the destruction of ferromagnetic order.

Our algorithm
enables us to investigate in detail how the ground states
change when ferromagnetic order is destroyed.  Barahona et
al.~\cite{BarahonaMar1982} present evidence that long range order in the
system is related to whether a set of spins with the same relative
orientation in {\em all} the ground states spans the system.  We can
refine this picture further by noting that the set of ground states can
naturally be subdivided into clusters~\cite{HartmannJan2001},
where a cluster is a group of ground
states related by a sequence of single spin flips, each of which leaves
the energy the same.  By definition, all states in the same cluster can
be reached from each other by single spin flips without raising the
energy, whereas ground states in different clusters can only be reached
from each other without raising the energy by making cooperative flips
of multiple spins.  A realization's ground states may all fall into a
single cluster or populate many distinct clusters.  For our systems, the
number of clusters is moderate---we have observed up to 12 clusters in a
single $10 \times 10$ realization.  It is natural to ask whether the
destruction of ferromagnetic order corresponds to growth of the number
of spins contributing to the individual clusters, or whether the
relationship between the clusters plays a vital role.

\begin{figure}
\centering
{\includegraphics[width=5cm]{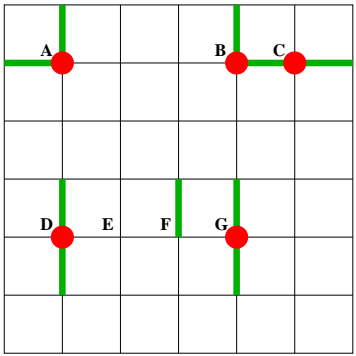}}
\caption{Three bunches of flippable spins.  The flippable spins are
denoted by filled-in circles and the frustrated bonds by thick lines.}
\label{fig:spinbunches}
\end{figure}

To address this question, it is useful to focus on the spatial relations
between the ground states in each cluster.  To do this, we first define
a flippable spin as one with an equal number of satisfied and
unsatisfied bonds; all the states in a cluster are related by sequential
flips of flippable spins.  We then define a bunch to be a group of
flippable spins in a given cluster whose flippability does not depend on
the state of other flippable spins in the system.
Figure~\ref{fig:spinbunches} shows three spin bunches.

The first bunch, spin $A$, is just a single isolated spin.  The second
bunch consists of spins $B$ and $C$.  Note that if spin $B$ is flipped,
then spin $C$ is no longer flippable, and conversely.  The third bunch
consists of spins $D$, $E$, $F$, and $G$.  The bunch contains all four
spins, because if $G$ is flipped, then $F$ becomes flippable.  If $D$
and $F$ are both then flipped, then $E$ becomes flippable.

Identification of bunches gives a complete picture of the clusters.
Different clusters cannot have all the same bunches, though a given
bunch can appear in more than one cluster.  Bunches are useful because
within a given cluster they are independent, so their contribution to
the ground state degeneracy is multiplicative.

\begin{figure}
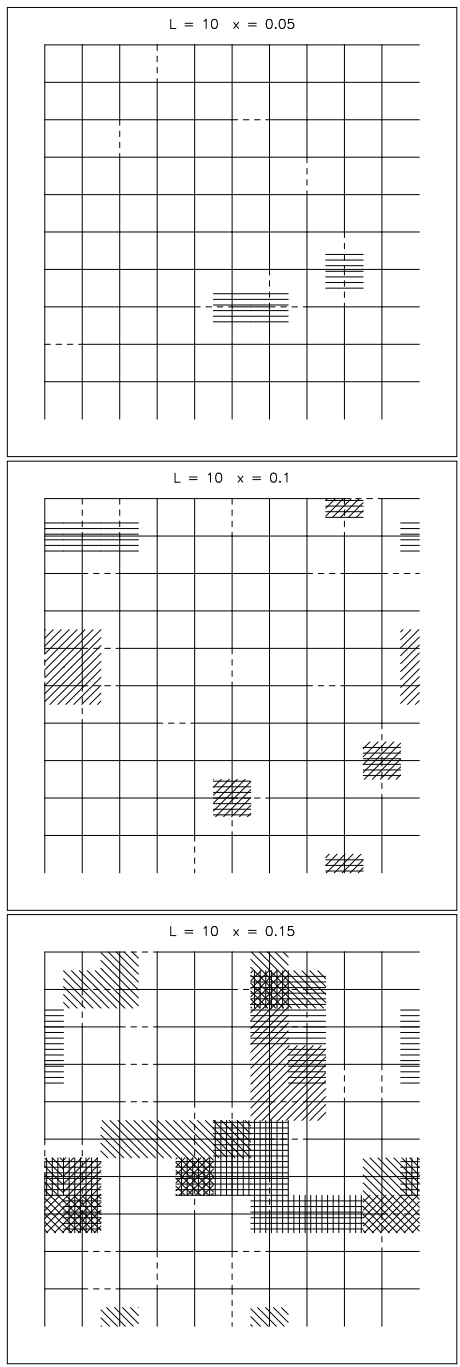

\centering
{\includegraphics[angle=270,width=6cm]{figures/10_10.180.bunch.latt.0-cluster.18a.ps}}
{\includegraphics[angle=270,width=6cm]{figures/10_20.6.bunch.latt.0-cluster.18b.ps}}
{\includegraphics[angle=270,width=6cm]{figures/10_30.46.bunch.latt.0-cluster.18c.ps}}
\caption{\label{fig:spanbunch}Three different $10 \times 10$ realizations at $x=0.05$,
$x=0.1 \approx x_c$, and $x=0.5$.  All the bunches in each cluster are
shown with hash marks, with different angles signifying different
clusters.  Multiple clusters allow bunches
to combine to span the space and destroy ferromagnetic order.}
\end{figure}

We extract from the complete set of ground states all the bunches of a
system using an algorithm described in Ref.~\cite{Landry2001}.
Figure~\ref{fig:spanbunch} shows bunches from three realizations with
$x=0.05 < x_c$, $x=0.1 \simeq x_c$, and one with $x=0.15 > x_c$.  One
can see that the bunch structure for a single cluster does not change
drastically as one crosses the transition, but when $x>x_c$ multiple
clusters exist and the overlap of all the bunches from the different
clusters spans the system.  We believe that the key element governing
the destruction of ferromagnetism is whether the overlap between the
different bunches in different clusters is such that the union of all
the bunches forms a path that percolates across the system.  Thus, the
``rigidity'' transition discussed by Barahona et
al.~\cite{BarahonaMar1982} is governed by overlap of the bunches
composing different clusters.

\begin{figure}
\centering
{\includegraphics[angle=270,width=8cm]{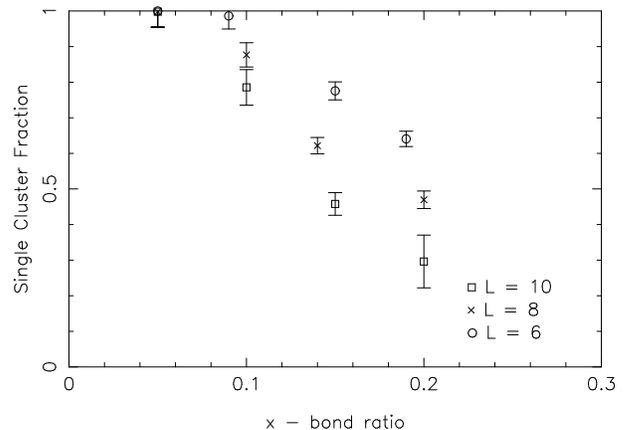}}
\caption{Plot of the single cluster fraction vs $x$ for $L=6$, $8$, and
$10$.  Error bars are estimated as $\sqrt{N_r}$, where $N_r$ is the number of
realizations. At $x=.05$, in the spin glass phase, the single cluster
fraction is essentially $1$.  As $x$ increases and passes through the
ferromagnetic transition, the single cluster fraction drops
precipitously.}
\label{fig:scfraction}
\end{figure}

Figure~\ref{fig:scfraction} shows the single cluster fraction as a
function of $x$ for $L = 6$, $8$, and $10$.  The number of realizations
used to generate this figure is shown in Table~\ref{tbl:scfraction}.
The presence of multiple clusters is strongly correlated with the
destruction of ferromagnetism.

\begin{table}
\centering
\begin{tabular}{|c|c|c|c|c|} \hline
$L$ & $x$ & $N_l$ & $N_r$ & $N_{scr}$ \\ \hline
6 & .0556 & 500 & 515 & 515 \\
6 & .0972 & 500 & 729 & 719 \\
6 & .1528 & 500 & 1212 & 940 \\
6 & .1944 & 500 & 1386 & 888 \\ \hline
8 & .0547 & 500 & 507 & 506 \\
8 & .1016 & 500 & 754 & 661 \\
8 & .1484 & 500 & 1187 & 738 \\
8 & .2031 & 281 & 775 & 364 \\ \hline
10 & .05 & 500 & 504 & 503 \\
10 & .1 & 200 & 317 & 249 \\
10 & .15 & 199 & 452 & 207 \\
10 & .2 & 18 & 54 & 16 \\ \hline
\end{tabular}
\caption{\label{tbl:scfraction}Table of the runs used to generate
Figure~\ref{fig:scfraction}.  $N_l$ is the number of distinct lattices,
$N_r$ is the number of ground state realizations, and $N_{scr}$ is the
number of ground state realizations with a single ground state cluster.}
\end{table}

We can investigate further the bunch overlap as $x$ is
increased through the spin glass transition.
We do this by defining an
overlap fraction for our realizations.
To do this, we first sum
up the number of spins in bunches for each cluster in the
realization, $n_{sb}$.
Then, all the bunches in all the clusters of a given
realization are overlaid, and the total number of spins covered by
bunches is
counted, $o_{sb}$.  We
define the overlap fraction $o_f$ as
\begin{equation}
o_f = \frac{n_{sb} - o_{sb}}{n_{sb}}
\end{equation}
If no spins overlap between the bunches in different clusters, $o_{sb} =
n_{sb}$ and $o_f = 0$.  If all the spins between different bunches in
$N$ different clusters overlapped (an impossibility), then $o_f$ would
tend toward $1$ as $N \rightarrow \infty$.  The overlap fraction $o_f$
is not defined for a realization with only a single cluster.

\begin{figure}
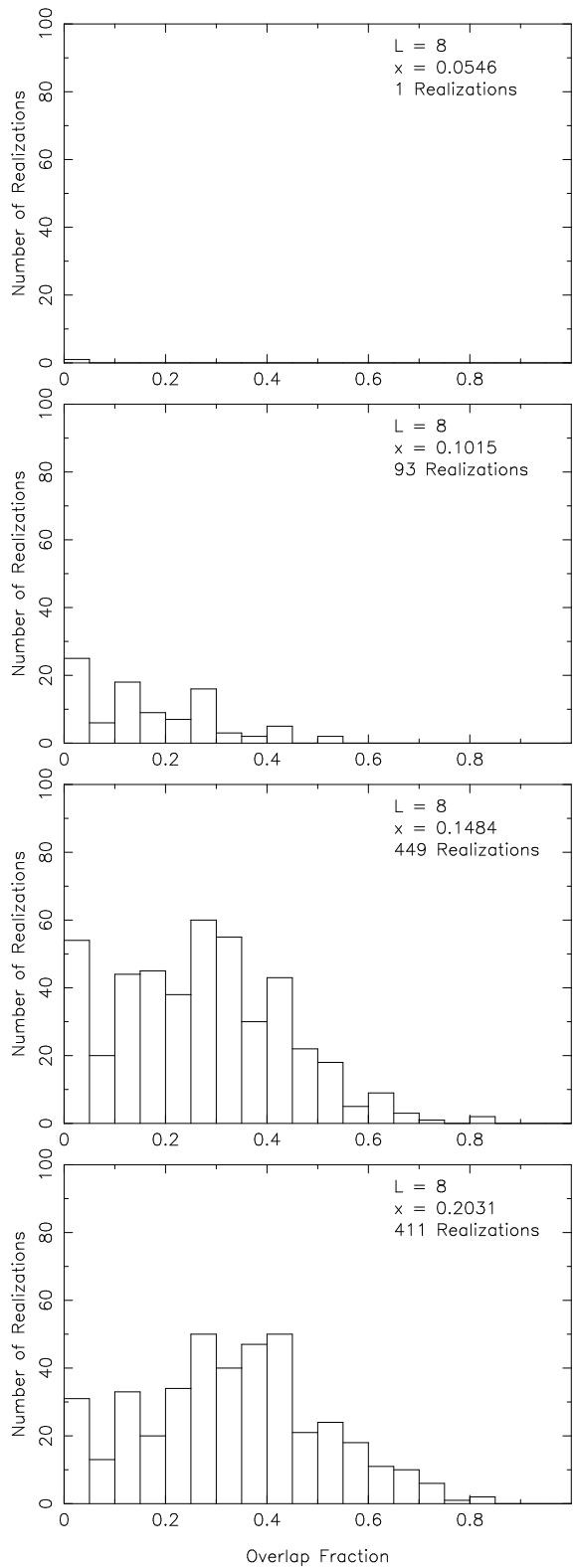

\centering
{\includegraphics[angle=270,width=7.5cm]{figures/8-7-500-overlap-hist.20a.ps}}
{\includegraphics[angle=270,width=7.5cm]{figures/8-13-500-overlap-hist.20b.ps}}
{\includegraphics[angle=270,width=7.5cm]{figures/8-19-500-overlap-hist.20c.ps}}
{\includegraphics[angle=270,width=7.5cm]{figures/8-26-281-overlap-hist.20d.ps}}
\caption{Plots of the overlap fraction $o_f$ for $L=8$ realizations.  As
the number of multiple cluster realizations increases, the structure of
the bunches also grows increasingly complicated as the overlap fraction
increases.}
\label{fig:overlap}
\end{figure}

Figure~\ref{fig:overlap}
shows histograms of $o_f$ for various $x$ values for lattices
with $L=8$.  As the number of
realizations with multiple clusters rises, the overlap fraction for
these clusters also increases.  Not only are there more clusters, but
the variability of the bunches in these clusters also increases.

Thus, the appearance of multiple clusters for a single realization and the
increase of overlaps between the bunches of these clusters both seem to be
closely correlated to the destruction of ferromagnetic order.  We
believe these additional markers of the transition will aid in the
understanding of this transition.

\section{\label{sec:discussion}Discussion}

This paper investigates the ground states of the
two-dimensional $\pm J$ Edwards-Anderson spin glass.  An algorithm that
finds systematically all the ground states of this model is presented.
This algorithm is used to investigate the nature of the destruction of
long-range ferromagnetic order as the fraction of antiferromagnetic
bonds is increased.

The running time of our algorithm for a given realization is found
empirically to scale roughly linearly in the number of ground states for
the range that we study (up to $10^7$ ground states).  Memory is the
most serious limitation, because to avoid duplication each new ground
state must be compared to those already found.  Our method enables the
examination of significantly larger systems than would be accessible by
exhaustive enumeration of all states --- for example, a system with
$L=10$ has $2^{100}\sim 10^{30}$ configurations, but only
$10^{2}-10^{7}$ ground states.  The advantage of our technique is even
greater when we introduce quantum tunneling~\cite{Landry2001}, where the
full Hamiltonian of a system with $L=10$ is a $2^{100} \times 2^{100}$
matrix.

We use our method to investigate the transition at which ferromagnetism
is destroyed as the fraction $x$ of antiferromagnetic bonds is
increased.  By comparing the ``rigidity'' of bonds between different
ground state configurations, we can obtain new insight into the
percolative nature of the phase transition at which ferromagnetism
disappears.

In general, we find that many quantities exhibit large, non-Gaussian
variability between realizations.  For example, the number of ground
states is well-described by a log-normal distribution, with the ratio of
the width to the mean of the distribution increasing with system size
$L$ up to the relatively large value $L \sim 8$.  This relatively large
length scale is unsettling when one keeps in mind the relatively small
sizes that can be investigated numerically
in higher dimensions~\cite{HartmannNov1999,HoudayerDec2000,%
KrzakalaOct2000,KatzgraberMay2001,Domany0104264}.

The availability of complete information about all the ground states
provides a new avenue for probing the nature of the system at low
temperatures, and our work can be extended in many directions.  We have
investigated how the introduction of quantum tunneling and of coupling
to a deformable lattice affect the low-energy landscape of the
model~\cite{Landry2001}, and the effects of other physical
perturbations should be examined.  Our work also provides a means for
stringent validation of other sampling methods that can be used to study
larger systems.  We also note that in the course of this work we have
shown that one can adapt the standard matching algorithm to study
systems with fully toroidal periodic and/or antiperiodic boundary
conditions.  It would be interesting to compare the effects of changing
boundary conditions of this type to those in
Ref.~\cite{Hartmann0107308}.

\begin{acknowledgments}
We gratefully acknowledge L. Saul for providing us with
the code of his program for finding the partition function
of a two-dimensional spin glass and J. Cook and A. Rohe
for putting the Concorde version of the blossom algorithm
in the public domain.  We also wish to thank Scot Shaw for
putting his ARPACK C++ routines in the public domain.

We thank J. Brooke, B. DiDonna, S.R. Nagel,
T.F. Rosenbaum, and S. Venkataramani for enlightening
conversations and J. Bogan for computer assistance.
This work was supported by MRSEC Program of the National
Science Foundation under Award Number
DMR-9808595.
\end{acknowledgments}

\appendix*

\section{Detailed Description of the Ground State Algorithm}
\label{sec:algorithm_details}

\subsection{Goals of the algorithm}

These notes are a short description of our algorithm that generates all
of the ground states of an Edwards-Anderson 2-D spin glass.  This
algorithm was designed with several goals in mind.

The algorithm should be relatively simple, so that it is easy to
understand and implement.

The algorithm should proceed in a completely deterministic and
systematic way, so there is no way to get off track or lose information.

The algorithm should be exhaustive.  Given infinite time and memory, it
should be able to find all the ground states of an arbitrary lattice.

The algorithm should use as much information as possible.  It should use
the fact that selecting one edge automatically eliminates many other
edges, simply because the two plaquettes joined by the edge cannot be
linked by any other edges.  It should also use the fact that the number
of edges in at least one ground state is a small subset of the total
number of edges.

\subsection{Short description}

The algorithm has two parts.  The first is preliminary and the second
uses the results of the first to generate all the ground states.

The first part finds all the viable edges: edges that exist in matching
solutions of the correct energy.  This list of viable edges is then the
only thing that needs to be studied, since the other edges do not exist
in matchings with the correct ground state energy.

The second part takes the list of viable edges and systematically goes
through the possible combinations (those that connect each plaquette to
one and only one other plaquette) and tests them to see if they
correspond to ground states.  Once all the combinations are exhausted,
all the ground states will have been found.

\subsection{Detailed outline of the algorithm}

\begin{enumerate}
\item Given $L$ and $x$, generate realization with randomly-placed bonds
\item Find the partition function of this realization
        \begin{enumerate}
        \item gives number of ground states
        \item gives energy of ground states
        \end{enumerate}
\item Convert this lattice to a matching problem in graph theory
        \begin{enumerate}
        \item frustrated plaquettes $\rightarrow$ nodes
        \item path of unsatisfied bonds between frustrated plaquettes
        $\rightarrow$ edges
        \end{enumerate}
\item Construct the list of viable edges (edges that appear in at least
        one ground state matching)
        \begin{enumerate}
        \item run blossom algorithm to find a matching solution of the
        correct energy.
        \item add the edges in this matching solution to the list of
        viable edges.
        \item build up the list of viable edges, one plaquette at a
        time, until all the edges corresponding to ground states are found.
                \begin{enumerate}
                \item remove all edges in the list for the current
        plaquette from the problem.     
                \item run blossom algorithm to get a new matching solution.
                \item if the new matching has the correct energy, add
        the new edges to the list of viable edges, return to i.
          and do it again.
                \item if it does not, there are no new edges for this
        plaquette.  Go to the next plaquette.
                \end{enumerate}
        \end{enumerate}
\item Take the list of viable edges and pair up the plaquettes until all
   ground states are found.
        \begin{enumerate}
        \item sequentially go through the various combinations of edges
      until every ground state is found.
                \begin{enumerate}
                \item pick series of edges until each plaquette is
                matched up with one and only one other plaquette.
                \item if matching solution does not have the correct
        energy, go to i.
                \item convert this matching solution to a set of spins
        and bonds for each boundary condition.  If this corresponds to a
        ground state, enter it for that boundary condition only.
                \item return to i. if some edge combinations remain.
                \end{enumerate}
        \item number of possible combinations is manageable,
      because each pick of an edge removes two plaquettes from the list
      that need to be matched.
        \end{enumerate}
\item When a matching solution is found, it does not necessarily
   correspond to a ground state, so all ground states are kept in
   memory.
        \begin{enumerate}
        \item one matching can represent several ground states. (A simple
      example is a plaquette diagonally separated by one from another 
      plaquette joined by an edge. This represents two ground states.)
        \item multiple matchings can represent one ground state. (Take the
      example of four plaquettes in a + formation linked together.  Many
      possible matchings represent the same ground state.)
        \item all ground states are kept in memory, and when a new
      matching solution is found, this is compared to ground states already
      found in the system.  If it is new, it is added to the list.
        \end{enumerate}
\item Once the algorithm is done, the number of ground states found is
   compared to the number found in step 2 as a check.
\end{enumerate}

\subsubsection{Finding all the viable edges}

This part of the algorithm uses removal of edges from the matching
problem (cutting) to find all the viable edges.  It works by repeatedly
cutting all the edges involving a specific plaquette until all matchings
generated are higher energy.  At that point, all viable matchings
contain one of the edges already cut.

\begin{figure}
\centerline{\includegraphics[height=6cm]{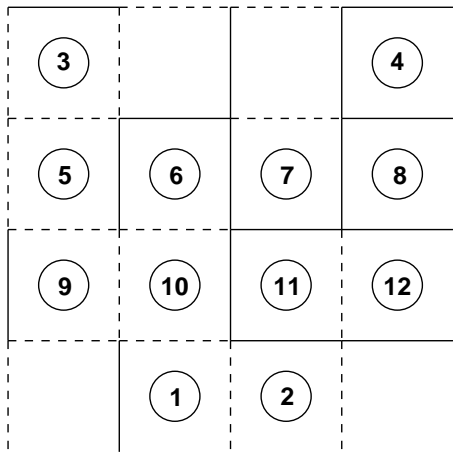}}
\caption{$4 \times 4$ Lattice for algorithm example.  Ferromagnetic bonds are dark
lines and antiferromagnetic bonds are dashed lines.  Frustrated
plaquettes are represented by open circles and labelled by number.}
\label{fig:alg-lattice}
\end{figure}

As an example, suppose we have a $4 \times 4$ lattice with $12$
plaquettes and $66$ edges.  The arrangement of bonds and frustrated
plaquettes is shown in Figure~\ref{fig:alg-lattice}.  

We run the blossom program without cutting any edges and get a ground
state matching with edges $(9,20,23,34,46,63)$.  Edge $9$ links
plaquettes $1$ and $10$, so we put edge $9$ in our list under both
plaquette $1$ and plaquette $10$.  We would do the same for the five
other edges in the matching.

We want to find all the edges that appear in ground state matchings that
connect to plaquette $1$.  We thus cut edge $9$ and feed the problem
back into the blossom algorithm.  We get the ground state matching
$(1,23,34,49,55,63)$.  Edge $1$ links plaquettes $1$ and $2$, so we put
edge $1$ in our list under both plaquette $1$ and plaquette $2$.  We do
the same for the other five edges.

We now cut both edge $9$ and edge $1$ and feed the result into the
blossom algorithm.  We get out the result $(8,20,23,34,49,56)$, a
matching with an energy larger than the ground state energy.  This means
that every ground state matching contains either edge $1$ or edge $9$.
We can then do the same for plaquette $2$, $3$, etc. until we have all
the edges that appear in ground state matchings, what we call viable
edges.

\subsubsection{Generating all the ground states}

Now that we have the list of viable edges, we restrict our attention to
it.  All ground states can be found by combining these edges.  We
combine these edges such every plaquette has one and only one other
plaquette connected to it by an edge.

As an example, consider the $4 \times 4$ example from above.

The set of possible edges, grouped by plaquette, looks like this:

\noindent plaquette 1: $(1,9)$\\
plaquette 2: $(1,20)$\\
plaquette 3: $(22,23)$\\
plaquette 4: $(22,34)$\\
plaquette 5: $(23,39,41,42)$\\
plaquette 6: $(39,46,49)$\\
plaquette 7: $(46,52,55)$\\
plaquette 8: $(34,41,52,60)$\\
plaquette 9: $(42,61,63)$\\
plaquette 10: $(9,49,61,64)$\\
plaquette 11: $(20,55,64,66)$\\
plaquette 12: $(60,63,66)$

We want to generate a matching with six edges.  Starting off is easy.
We merely pick the first one in plaquette 1's list, edge $1$.  Edge $1$
involves plaquettes $1$ and $2$, so we skip to plaquette $3$'s list and
select edge $22$.  That involves edges $3$ and $4$, so the next highest
plaquette is $5$.  We select edge $23$, but that involves plaquette $3$
and is forbidden, so we skip to the next one, which is $39$.  This
continues until all the plaquettes are accounted for and we get the
matching $(1,22,39,52,61,66)$.  We would then convert this matching into
possible ground states for each boundary condition.  Those potential
ground states that have consistent spin configurations and energes are
entered into the ground state lists.

Now we go to the next state.  We remove the last two edges from our
current state, giving us $(1,22,39,52)$.  The second to last plaquette
list we looked at was $9$, so we choose the edge after edge $61$.  We
get $63$, which involves plaquettes $9$ and $12$.  We then skip to
plaquette $10$ and choose edge $9$.  That involves plaquette $1$, which
is already linked.  Edge $49$ involves plaquette $6$ which is taken, and
edge $61$ involves plaquette $9$ which is already taken.  Only edge $64$
involves the empty plaquettes $10$ and $11$, so we get the new state
$(1,22,39,52,63,64)$.

We continue in this manner to get all of the ground states.  If we reach
the end of a list for a specific plaquette, we know to erase that entry
in the current state and move to the previous plaquette list.  If we
reach the end of the list for plaquette $1$, we know we have completed
the algorithm and found all the ground states.

\subsection{Notes on the algorithm}

The algorithm currently is memory-limited.  The blossom algorithm and
selection of edges for each matching in part 5 are both very fast.
Eventually, as the number of ground states grows with the size of the
system and the value of $x$, the available memory of the computer is
exhausted. 

The reason every ground state is stored in memory now is that in certain
circumstances, distinct matching solutions can correspond to the same
ground state.  If these circumstances were determined and recorded, then
only a small subset of relevant ground states would have to be kept,
while the others are written to a file.  Whenever a matching solution
with a ``dangerous'' edge or edges came up, it could be compared to the
small subset of comparable matching solutions or ground states.  It is
not clear if this approach is really feasible or not.

Finally, certain combinations of edges that automatically lead to higher
energy matching solutions constantly recur.  If a list of these could be
quickly generated, it should greatly decrease the number of combinations
of viable edges that must be explored.

\bibliography{sg}
\end{document}